\documentclass[prd,twocolumn,preprintnumbers,reprint,amsmath,amssymb]{revtex4-1}

\usepackage{graphicx}
\usepackage[tight]{subfigure}

\newcommand\fverb{\setbox\pippobox=\hbox\bgroup\verb}
\newcommand\fverbdo{\egroup\medskip\noindent%
                        \fbox{\unhbox\pippobox}\ }
\newcommand\fverbit{\egroup\item[\fbox{\unhbox\pippobox}]}

\newcommand{\im}{\mathrm{Im}}

\newcommand{\colorparbox}[3]{\colorbox{#1}{\parbox{#2}{#3}}}
\newcommand{\colorwideparbox}[2]{\colorparbox{#1}{\textwidth-5mm}{#2}}

\newcommand{\chapter}[1]{\color{fgSection}\colorwideparbox{bgSection}{%
    \centering\bfseries\boldmath#1}}

\newcommand{\tb}{\tan\!\beta}
\newcommand{\bsg}{B \rightarrow X_s \gamma}
\newcommand{\dms}{\Delta M_s}
\newcommand{\bsbsbar}{$B_s$--$\overline{B_s}$}

\newcommand{\phis}{\phi_s}
\newcommand{\phiq}{\phi_q}
\newcommand{\Absl}{A^b_\mathrm{sl}}
\newcommand{\adsl}{a^d_\mathrm{sl}}
\newcommand{\assl}{a^s_\mathrm{sl}}
\newcommand{\aqsl}{a^q_\mathrm{sl}}

\newcommand{\deltad}{\delta^d}
\newcommand{\ded}[2]{(\deltad_{#1})_{#2}}

\begin{document}

\preprint{DESY 10--097}

\title{Addendum to: Implications of the measurements of
{\boldmath\bsbsbar} mixing on SUSY models}

\author{P.~Ko}
\affiliation{School of Physics, KIAS, Seoul 130--722, Korea}

\author{Jae-hyeon Park}
\affiliation{Deutsches Elektronen-Synchrotron DESY,
  Notkestra{\ss}e 85, 22603 Hamburg, Germany}

\begin{abstract}
  This is an addendum to the previous publication,
  P.~Ko and J.-h.~Park, Phys.\ Rev.\ {\bf D80}, 035019 (2009).
  The semileptonic charge asymmetry in $B_s$ decays is discussed
  in the context of general MSSM with gluino-mediated flavor 
  and CP violation in light of the recent measurements
  at the Tevatron.
\end{abstract}
\maketitle

In this addendum to Ref.~\cite{Ko:2008xb}, 
we discuss the semileptonic charge asymmetry
in the $B_s$ decays in general SUSY models with gluino-mediated 
flavor and CP violation, in light of the recent measurements of
like-sign dimuon charge asymmetry by D\O\ Collaboration 
at the Tevatron.  The model is described in Ref.~\cite{Ko:2008xb}, to which we refer for the details of the model and other 
phenomenological aspects related with $B_s - \overline{B_s}$
mixing, the branching ratio of and CP asymmetry in 
$B\rightarrow X_s \gamma$, $B_d \rightarrow \phi K_S$ and 
CP asymmetry in $B_s \rightarrow J/\psi \phi$. 

One can define the semileptonic charge asymmetry
in the decay of $B_q$ mesons as
\begin{equation}
  \label{eq:def ASL}
  \aqsl \equiv
  \frac{\Gamma(\overline{B^0_q}(t) \rightarrow \mu^+ X) -
    \Gamma(B^0_q(t) \rightarrow \mu^- X)}%
  {\Gamma(\overline{B^0_q}(t) \rightarrow \mu^+ X) +
    \Gamma(B^0_q(t) \rightarrow \mu^- X)}
  ,
\end{equation}
for $q = d, s$.
In terms of the matrix elements of the effective Hamiltonian describing
the damped oscillation between $B^0_q$ and $\overline{B^0_q}$,
the asymmetry $\aqsl$ is given by
\begin{equation}
  \label{eq:ASL phis}
  \aqsl = \im \frac{\Gamma^q_{12}}{M^q_{12}}
  = \frac{|\Gamma^q_{12}|}{|M^q_{12}|} \sin\phiq
  ,
\end{equation}
where $\phiq \equiv \arg(-M^q_{12}/\Gamma^q_{12})$.
That is, this is another observable measuring $CP$ violation
in $B_q$--$\overline{B_q}$ mixing.
We take the approximation,
$\Gamma^q_{12} = \Gamma^{q,\mathrm{SM}}_{12}$,
since the leading contribution comes from the absorptive part of
the box diagrams for $B_q$--$\overline{B_q}$ mixing and
there is no new common final state into which
both $B_q$ and $\overline{B_q}$ can decay in our scenario.
The size of $M^q_{12}$ is fixed by the $\Delta M_q$ data
up to hadronic uncertainties.
Then, $\aqsl$ can be regarded as a sine function of $\phi_q$, multiplied by
the factor $|\Gamma^q_{12}|/|M^q_{12}|$.
This curve is traversed as one allows for arbitrary supersymmetric
contributions to $M^q_{12}$ obeying the $\Delta M_q$ constraint.
Combining the SM predictions \cite{Lenz:2006hd},
\begin{equation}
  \label{eq:G12SM}
  \begin{aligned}
|\Gamma^{s,\mathrm{SM}}_{12}|/|M^{s,\mathrm{SM}}_{12}| &=
(49.7\pm 9.4)\times 10^{-4} ,
\\
\phis^\mathrm{SM} &= (4.2 \pm 1.4)\times 10^{-3} ,
  \end{aligned}
\end{equation}
one finds the vanishingly small asymmetry
$a^{s,\mathrm{SM}}_\mathrm{sl} \sim 2\times 10^{-5}$.

Recently, the D\O\ collaboration reported a measurement of
like-sign dimuon charge asymmetry \cite{Abazov:2010hv}.
They interpreted the result as coming from the mixing of
neutral $B$ mesons and have found an evidence for an anomaly
in the asymmetry,
\begin{equation}
  \Absl \equiv \frac{N_b^{++} - N_b^{--}}{N_b^{++} + N_b^{--}}
  ,
\end{equation}
where $N_b^{++}$ and $N_b^{--}$
are the number of events where
decays of two $b$ hadrons yield
two positive and two negative muons, respectively.
Their result shows a discrepancy of $3.2 \sigma$ from the SM expectation.
This asymmetry consists of $\adsl$ coming from $B_d$ decays
as well as $\assl$ from $B_s$.
One can extract the asymmetry relevant to the $B_s$ meson
using the measured value of $\adsl$ and the result by D\O\ is
\begin{equation}
  \assl = -0.0146 \pm 0.0075.
\end{equation}
This is $1.9 \sigma$ away from the SM prediction.
We shall use this data in the following discussion.

This D\O\ result has drawn interest in new physics explanations
\cite{Ligeti:2010ia,Dighe:2010nj,Bauer:2010dg,NP in M12,Parry:2010ce}.
(For earlier works,
see e.g.\ Refs.~\cite{Kane et al,Lenz:2007nj,All in NP}.)
Some of the works consider extra contributions to $\Gamma^q_{12}$
since the dimuon charge asymmetry depends on it as well
as on $M^q_{12}$ \cite{Dighe:2010nj,Bauer:2010dg}.
This approach also has a possibility of altering $|\Delta \Gamma_s|$
even though its current experimental value is in agreement with
the SM one, $2\, |\Gamma^{s,\mathrm{SM}}_{12}\cos\phis^\mathrm{SM}|$
\cite{Lenz:2006hd,Barberio:2008fa,Esen:2010jq}.
As we said, $\Gamma^q_{12}$ is fixed in the present work
and we are left only with the option of modifying $M^s_{12}$.
Therefore, $|\Delta \Gamma_s|$ shall become smaller than its SM prediction
as $|\phis|$ grows up to $\mathcal{O}(1)$.

We perform the numerical analysis in the same way as in the main article
\cite{Ko:2008xb}.
The crucial ingredient for evaluating $\assl$ is
the range of $\phis$ to be used.
Following the latest reports from D\O\ \cite{Abazov:2010hv} and CDF
\cite{CDF FPCP2010},
there have been a couple of attempts to make a global fit
of \bsbsbar\ mixing parameters including $\phis$
\cite{Ligeti:2010ia,Bauer:2010dg}.
However, the official combination is not available yet.
Partly because of this reason and partly for the sake of
coherent presentation,
we keep using the range used in Refs.~\cite{Ko:2008xb,Bona:2008jn},
\begin{equation}
  \phis \in [-1.10,-0.36] \cup [-2.77,-2.07] .
\end{equation}
As a matter of fact,
this range is not very different from the $2 \sigma$ interval found in
Ref.~\cite{Bauer:2010dg}.
As for $\Gamma^{s,\mathrm{SM}}_{12}/M^{s,\mathrm{SM}}_{12}$,
we take its central value from Eqs.~(\ref{eq:G12SM}).
Considering the error in this ratio could add
20\% more of uncertainty to the thickness of
the $\assl$ band in the following figures.

\begin{figure*}
  \centering
  \subfigure[\ $LL$\hspace{-4em}]%
  {\includegraphics[height=62mm]{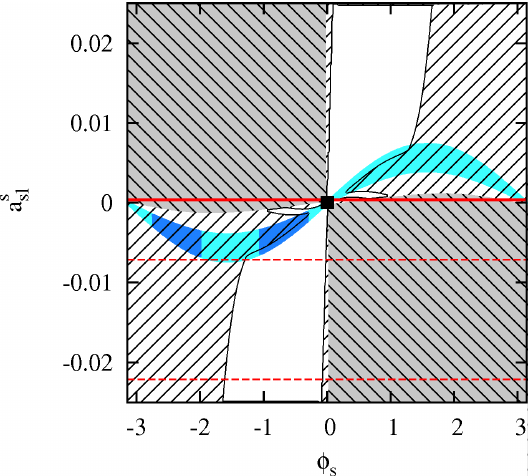}}\qquad
  \subfigure[\ $RR$\hspace{-4em}]%
  {\includegraphics[height=62mm]{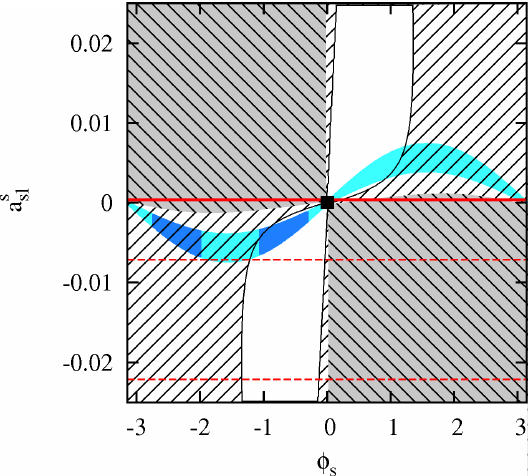}}
  \\
  \subfigure[\ $LL=RR$\hspace{-4em}]%
  {\includegraphics[height=62mm]{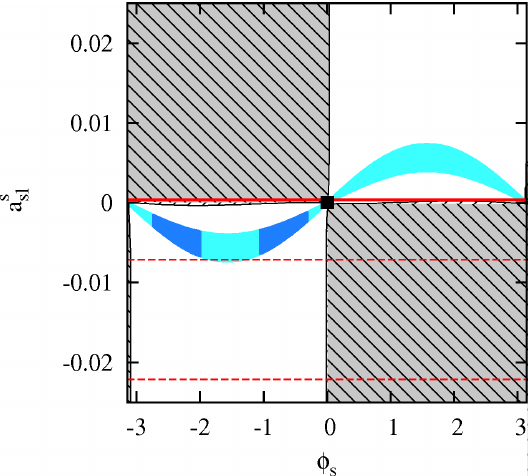}}\qquad
  \subfigure[\ $LL=-RR$\hspace{-4em}]%
  {\includegraphics[height=62mm]{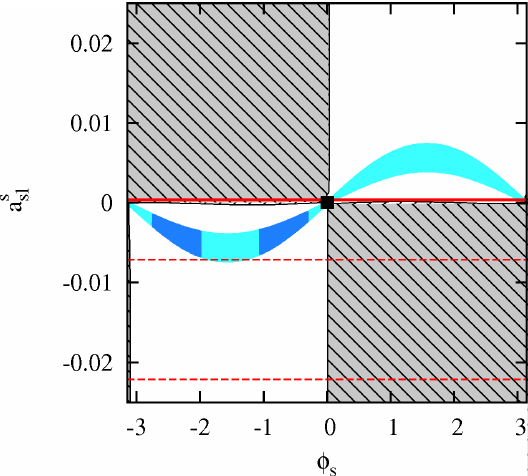}}
  \caption{Plots of $\assl$ as a function of $\phis$
    for the four different cases with $\tb = 3$.
    The hatched gray region leads to the lightest squark mass $< 100$ GeV.
The hatched region is excluded by the $B\rightarrow X_s \gamma$ constraint.
The light gray region (cyan online) is allowed by $\dms$.
The dark gray region (blue online) is allowed both by $\Delta M_s$ and $\phis$.
The black square is the SM point.
The dashed and solid lines (both red online) mark the
$1\sigma$ and $2\sigma$ ranges of $\assl$, respectively.}
  \label{fig:ASL tanb=03}
\end{figure*}
We show $\assl$ as a function of $\phis$
for $\tb = 3$ in Figs.~\ref{fig:ASL tanb=03}.
The four plots are for the $LL$, the $RR$, the $LL=RR$,
and the $LL=-RR$ cases, respectively.
One can immediately notice the aforementioned sinusoidal dependence of
$\assl$ on $\phis$, coming from Eq.~(\ref{eq:ASL phis})
and the $\dms$ constraint.
This feature is not only true of all the cases shown here
but also of any new physics model that does not affect $\Gamma^s_{12}$.
The nonzero thickness of the band arises from the uncertainty in $\dms$.
The difference between $\assl$ and its central value
is at least about $1.0 \sigma$.
This discrepancy becomes worse but only slightly
after $\phis$ is restricted inside its preferred ranges (colored in blue).
If one incorporates the $\bsg$ constraint, substantial part of
the blue regions is excluded,
in particular in the upper two cases with one insertion.
Even then, however, the lowest possible value of
$\assl \simeq -0.006$ within the blue region does not change.
In the lower two cases with two insertions,
$\bsg$ does not play an important role since
the supersymmetric effect on \bsbsbar\ mixing is enhanced.

\begin{figure*}
  \centering
  \subfigure[\ $LL$\hspace{-4em}]%
  {\includegraphics[height=62mm]{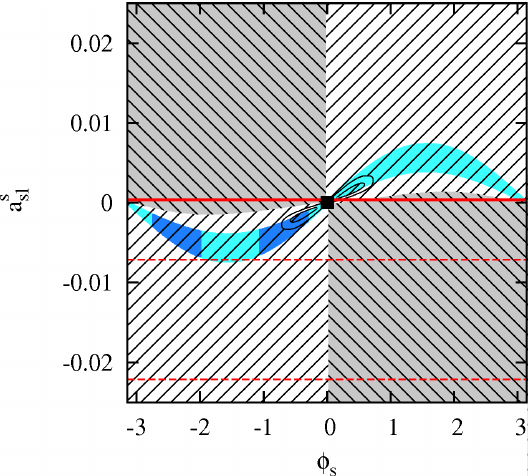}}\qquad
  \subfigure[\ $RR$\hspace{-4em}]%
  {\includegraphics[height=62mm]{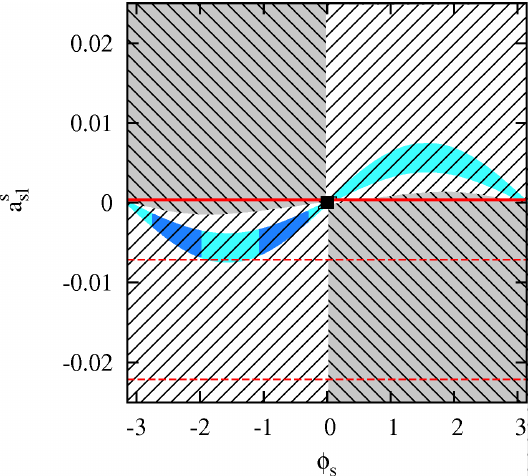}}
  \\
  \subfigure[\ $LL=RR$\hspace{-4em}]%
  {\includegraphics[height=62mm]{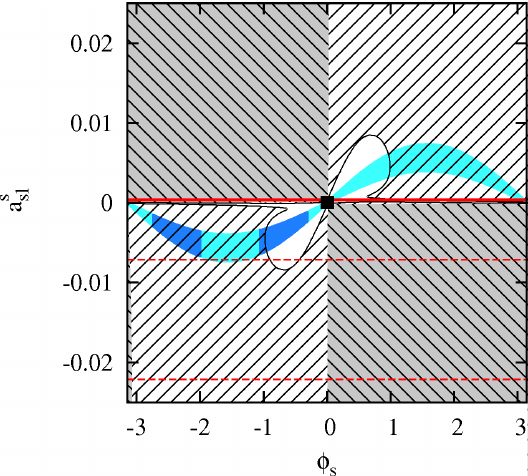}}\qquad
  \subfigure[\ $LL=-RR$\hspace{-4em}]%
  {\includegraphics[height=62mm]{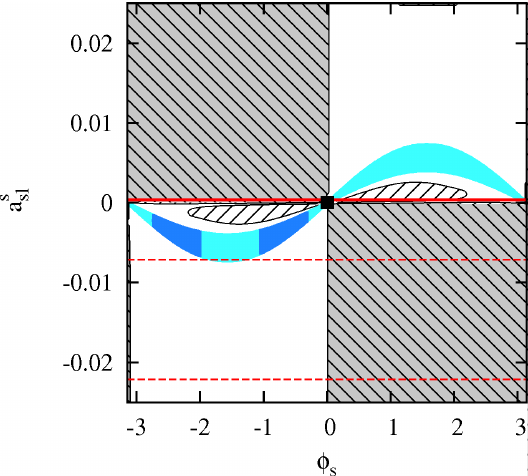}}
  \caption{Plots with $\tb = 10$.
    The meaning of each region is the same as in
    Figs.~\ref{fig:ASL tanb=03}.}
  \label{fig:ASL tanb=10}
\end{figure*}
Plots for $\tb = 10$ are displayed in Figs.~\ref{fig:ASL tanb=10}.
The model-independent characteristics
dictated by Eq.~(\ref{eq:ASL phis}) remain exactly the same
as in the previous set of figures.
The only difference is the stronger $\bsg$ constraint
due to higher $\tb$.
Here, it excludes more part of the blue regions.
Again, this is particularly true of the upper two cases in which
$\assl$ is restricted closer to its SM value.
In Fig.~\ref{fig:ASL tanb=10}(a),
$\dms$, $\phis$, and $\bsg$, together allow
$\assl$ to be as low as $-0.003$.
In Fig.~\ref{fig:ASL tanb=10}(b),
there is no solution satisfying all the three constraints.
One could get $\assl \simeq -0.0006$ if $\phis$ were not limited.
In the lower two cases, the lowest $\assl$,
compatible with $\dms$ and $\phis$, is almost the same as in
Figs.~\ref{fig:ASL tanb=03}.

We summarize.
We have examined how $\assl$ is influenced by
the $LL$ and/or $RR$ mass insertions.
For $\tb = 3$, one can reduce
the discrepancy between $\assl$ and its SM expectation
from $1.9 \sigma$ down to $1.0 \sigma$
in each of the $LL$, $RR$, $LL=RR$, and $LL=-RR$ cases,
obeying the $\dms$, $\bsg$, and $\phis$ constraints.
This amounts to reduction of the $\Absl$ tension from
$3.2 \sigma$ down to $2.2 \sigma$ if one assumes no new physics
in the $b \rightarrow d$ transition.
For $\tb = 10$, it becomes difficult for the $LL$ and $RR$ cases
whereas the $LL=RR$ and $LL=-RR$ cases are less limited by $\bsg$.

\begin{acknowledgments}
We thank Ahmed Ali, Alexander Lenz, and Satoshi Mishima for useful comments.
\end{acknowledgments}

\section*{Note added}

While we were waiting for the approval for submission,
a paper by J.~K.~Parry appeared on the e-print archive
that employs a related model \cite{Parry:2010ce}.
However, the flavor structure of the squark mass matrix therein
is different from any of those here.
As far as squarks are concerned,
he considers only one case where
$\ded{23}{RR}$ is a variable parameter
and $\ded{23}{LL}$ is fixed to a value
that comes from renormalization group running.
This way of parameter scan is not covered in this work.
He does not display the $\bsg$ constraint
on his plots, but it may not be very restrictive in his case
depending on $\mu$ and $\tb$.
(See e.g.\ Fig.~4 in Ref.~\cite{Ko:2008zu}.)

\end{document}